\documentclass[10pt]{article}
\usepackage{newinutile-2.0}
\usepackage{cite}

\usepackage{latexsym,amsmath,amsfonts,amsthm,amssymb,bbm}
\usepackage{epsfig,graphics,color,calc,graphicx,pict2e}
\usepackage{comment} 
\usepackage{multirow}
\usepackage{footmisc} 
\usepackage[T1]{fontenc}
\usepackage[utf8]{inputenc}
\usepackage{multicol}
\usepackage{float} 

\usepackage{mathrsfs}
\usepackage{tikz}
\usepackage{tikz}
\usetikzlibrary{shapes,arrows}
\usetikzlibrary{intersections,matrix, positioning}

\usepackage{todonotes}
\usepackage[printonlyused,withpage]{acronym}
\usepackage[acronym]{glossaries}

\usepackage{subfigure} 

\newcommand{\newatop}[2]{\genfrac{}{}{0pt}{}{#1}{#2}}

\newlength{\pecettawidth}
\makeglossaries
\newacronym{pca}{PCA}{Probabilistic Cellular Automata}
\newacronym{ca}{CA}{Cellular Automata}
\newacronym{eca}{ECA}{Elementary Cellular Automata}
\newacronym{deca}{DECA}{Diploid Elementary Cellular Automata}
\newacronym{ndeca}{NDECA}{Null Diploid Elementary Cellular Automata}

\begin{document}
\title{\normalsize\Large\bfseries Block approximations for probabilistic mixtures of elementary cellular automata}


\author[1]{Emilio N.M. Cirillo\thanks{emilio.cirillo@uniroma1.it}}
\affil[1]{Dipartimento di Scienze di Base e Applicate per l'Ingegneria, 
             Sapienza Universit\`a di Roma, 
             via A.\ Scarpa 16, I--00161, Roma, Italy.}

\author[1]{Giacomo Lancia\thanks{giacomo.lancia@uniroma1.it}}

\author[2]{Cristian Spitoni\thanks{C.Spitoni@uu.nl}}
\affil[3]{Institute of Mathematics,
University of Utrecht, Budapestlaan 6, 3584 CD Utrecht, The~Netherlands.}

\date{\empty} 

\maketitle

\begin{abstract}
Probabilistic Cellular Automata are a generalization of Cellular Automata. 
Despite their simple definition, they exhibit fascinating and 
complex behaviours. The stationary behaviour of these models 
changes when model parameters are varied, making the study of their 
phase diagrams particularly interesting. The block approximation method, 
also known in this context as the local structure approach, is a powerful 
tool for studying the main features of these diagrams, improving 
upon Mean Field results. This work considers systems with multiple stationary states, aiming to understand how their 
interactions give rise to the structure of the phase diagram.
Additionally, it shows how a simple algorithmic 
implementation of the block approximation allows for the effective study 
of the phase diagram even in the presence of several absorbing states.
\end{abstract}






\begin{multicols}{2}



\section{Introduction}
\label{s:intro} 
\par\noindent
\gls{pca} generalize deterministic \gls{ca} as a 
discrete--time Markov chains with state space obtained 
as the product of a single lattice site finite state space.
The variables associated with 
the lattice sites are updated independently and simultaneously according with 
a single site 
probability distribution which depends on the configuration 
of a neighbourhood of the site at the previous time.
Despite the simplicity of their stochastic evolution rules,  
\gls{pca} exhibit a large variety of dynamical behaviours and, 
for this reason, they are considered powerful modelling tools,
see, e.g., \cite{LN2016} for a general introduction to the topic
and \cite{VRD2023,D2020} for recent biological applications.

In this paper, we study the stationary behaviour of a particular 
class of \gls{pca} which are a probabilistic mixtures of 
\gls{eca} \cite{W83,W84}, namely, one dimensional \gls{ca} in which 
the state of a cell is described by a two--state variable whose actual
value, either zero or one, depends on the previous state of the two neighbouring 
cells and on that of the cell itself. 
In this context, 
binary
probabilistic 
mixtures, called \emph{diploid}, have been widely studied 
in the recent literature \cite{Fautomata2017}, not only for their intrinsic
interest, but also because some of them encode the evolution 
rule of very well known paradigmatic \gls{pca} models, such as 
the \emph{percolation} \cite{BMM2013,T2004}, 
the  \emph{noisy additive} \cite{MM2014}, 
the \emph{Stavskaya} \cite{Me2011},
and the \emph{directed animals}  \gls{pca} \cite{Dh83}.

We will use block approximation techniques as the main tool of our analysis. 
The idea of approximating the stationary measure of the 
\gls{pca} via measures defined on blocks with a maximal 
controlled size dates back to the 
pioneering papers \cite{GVK87,GV87}
in which the local structure of \gls{pca} was
firstly investigated.
More recently, this idea has been used to investigate the structure 
of the stationary measures of \gls{pca} in several contexts, 
for instance, in the case of asynchronous \gls{pca}
\cite{FF15}.

In these applications, the block approximation is often 
implemented by writing analytic expressions for some 
quantity of interest \cite{F12,Me2011a}.
In this work, we adopt an algorithmic scheme inspired 
by the Cluster Variation Method \cite{P05,CGP1996}, which is commonly used for 
Statistical Mechanics systems.
We emphasize that we do not look for optimization 
on the number of variables or for the analytic control of 
the computation, but we choose the maximal block and write 
recursive equations for the probabilities associated with each possible 
configuration on that block. 
The advantage is that we set a fully algorithmic procedure 
which, in principle, can be applied to any \gls{pca}.
On the negative hand we, obviously, miss any analytical control 
on the computation which, additionally, becomes more and more time--consuming 
when the size of the maximal block is increased.

The aim of this paper is to show the ability of this basic 
implementation of the block approximation
to describe the structure of the stationary measures 
of \gls{pca} in several contexts, in particular, in 
cases in which competition between contrasting behaviours 
makes the model complicated and not 
accessible by simple Mean Field techniques. 
More precisely, the goal is to use the block approximation 
to construct the phase diagram of the model, that is to say, 
to characterize the stationary measure of the model in the 
different regions of the parameter space and, in particular, 
to detect those points (or lines) where a transition between different stationary states is observed.

As the first goal, we consider the diploid, in which one of the 
two local rules is the null rule, namely the rule associating the zero state with any local configuration. 
These models will be called \gls{ndeca}.
Due to its interesting behaviour, this particular class of diploid 
has been extensively studied in \cite{Fautomata2017} via Monte 
Carlo simulations and in \cite{CNS2021} via 
Mean Field and rigorous one--site ergodicity criteria
\cite{MScmp1991,MScmp1993}.
The Mean Field analysis is in general in very good agreement 
with the Monte Carlo results, but for some specific 
choices of the non--null rule, the two methods are 
in disagreement on the existence of a transition or on its 
character. 
Here, we show that these discrepancies are resolved 
when approximations on blocks larger than three are considered. 
Moreover, looking at the block probability distribution, 
it is possible to understand 
why the Mean Field approximation fails
in describing the diploid behaviour. 

We remark that one of the characterizing traits of 
\gls{ndeca} is that there exists at most one unique absorbing state, namely, the zero state, since cells with state one have a non--zero probability to be changed to zero. 
For this reason,
in the second part of the paper
we focus our attention on models in which, on the contrary,
there exist at least 
two absorbing states.
We will, thus, analyze how their contrasting effects 
affect the stationary behavior of the \gls{pca} and we will
test the ability of our implementation of the block 
approximation to detect the structure of the phase diagram 
even in this intrinsically pernicious set--up.

We have chosen two models such 
that one of them can be 
written as a diploid, namely, a probabilistic mixture of 
two \gls{eca}, while the second one is a probabilistic 
mixture of four \gls{eca}.
This second model is equivalent to an example 
of totalistic \gls{pca}, namely, a \gls{pca} 
in which the site updating probability depends 
only on the total of the values of the cells in a neighbourhood. 
In both cases, we show that the block 
approximation is able to provide a complete interpretation of the 
Monte Carlo results and to give a full description of the main 
features of the model. 
The two selected models are not new in the literature, and
their interest 
have already been pointed out in 
\cite{Me2011a,Ba2000}.
 
The paper is organized as follows. 
We give the definition of a probabilistic mixture of 
\gls{eca} and introduce the block approximation in 
Section~\ref{s:modello}. 
We discuss the \gls{ndeca} in Section~\ref{s:ndeca},
while the diploid and the four \gls{eca} mixture 
with two absorbing states 
are respectively discussed in Sections~\ref{s:deca_182_200} and 
\ref{s:1d_bbr}.
Finally, in Section~\ref{s:con} we summarize our conclusions.

\section{Model and methods}
\label{s:modello} 
\par\noindent
In this section, we first introduce the class of models which will be the 
object of our study and then we will discuss our main techniques. 

\subsection{Probabilistic mixtures of \gls{eca}}
\label{s:mixture} 
\par\noindent
Given
the \emph{set of states} 
$Q=\{0,1\}$ 
and 
the
$n$ cell
annulus $\Lambda_n=\mathbb{Z}/n\mathbb{Z}=\{0,1,\dots,n-1\}$, 
with $n$ an odd integer number, 
we consider 
the \emph{configuration space}
$X_n=Q^{\Lambda_n}$.
For $x\in X_n$, $x_i$ is called the \emph{value} of 
the cell $i$ or the \emph{occupation number} of the cell $i$. 
The configurations with all the cell states equal to 
zero (resp.\ one) will be simply denoted by $0$ (resp.\ $1$).
For any $x\in X_n$ and $\Delta\subset\Lambda_n$, we 
let $x_\Delta$ be the restriction of $x$ to $\Delta$. 

In \gls{ca} the cell states 
are updated synchronously according to the state 
the cells of the system. 
More precisely, 
given the map $F:X_n\to X_n$,
the \gls{ca} associated with $F$ 
is the 
collection of all the sequences of configurations 
$(\zeta^t)_{t\in\mathbb{N}}$ obtained by applying the 
map $F$ iteratively, namely, such that $\zeta^t=F(\zeta^{t-1})$.
The sequence $(\zeta^t)_{t\in\mathbb{N}}$ such that $\zeta^0=x\in X_n$ 
is called \emph{trajectory} of the \gls{ca} with \emph{initial condition} $x$.

A particular example of \gls{ca} are the so--called 
\gls{eca}, in which the cell states 
are updated according to the state 
of the two neighbouring cells and that of the cell itself.
These three cells form the neighbourhood of the cell, namely, 
the \emph{neighborhood of the cell} $i\in\Lambda_n$ is 
the set $I_i=\{i-1,i,i+1\}$.
Note that, due to the periodic structure of the lattice $\Lambda_n$, 
the neighbourhood of the origin is the set 
$I_0=\{n-1,0,1\}$.
Thus, given a \emph{local rule} $f:Q^3\to Q$,
the associated \gls{eca} is the \gls{ca} defined by the 
map
$F:X_n\to X_n$ such that
$(F(x))_i = f(x_{I_i}),$ for any $i\in\Lambda_n$.

We recall the classical notation for all the possible 256 \gls{eca}, 
see, e.g., \cite{CNS2021,Fautomata2017}
and references therein:
a local rule $f$ is identified by the integer 
number $W\in\{0,\dots,255\}$ such that 
\begin{align}
\label{fin000}
W
=&
\phantom{+}
f(1,1,1)\cdot2^7
+
f(1,1,0)\cdot2^6
+
f(1,0,1)\cdot2^5
\notag\\
&
+
f(1,0,0)\cdot2^4
+
f(0,1,1)\cdot2^3
+
f(0,1,0)\cdot2^2
\notag\\
&
+
f(0,0,1)\cdot2^1
+
f(0,0,0)\cdot2^0
\notag\\
=
&\sum_{i=0}^7 c_i\cdot 2^i
.
\end{align}
The last equality defines the coefficients $c_i$ providing 
the binary representation of the number $W$. 
We shall sometimes denote the \gls{eca} 
with both the decimal and the binary representation by writing 
$W(c_7c_6c_5c_4c_3c_2c_1c_0)$.
Note that for the given the \gls{eca} $W$ as in \eqref{fin000}, the 
\textit{conjugate under left--right reflection} 
rule is obtained by exchanging $f(1,1,0)$ with $f(0,1,1)$ 
and $f(1,0,0)$ with and $f(0,0,1)$.

The stochastic generalization of \gls{ca} are called 
\gls{pca} and are defined as a Markov Chain 
$(\zeta^t)_{t\in\mathbb{N}}$ on the configuration space $X_n$ 
with transition matrix 
\begin{equation}
\label{mod000a}
p(x,y)
=
\prod_{i\in\Lambda_n}
p_i(y_i|x),
\end{equation}
where $p_i(\cdot|x)$, called \emph{one--site updating probability},
is a probability distribution on $Q$ parameterized 
by the configuration $x\in X_n$. 
Thus, at each time all the cells are updated 
simultaneously and independently. 
We denote by $P_x$ the probability associated with the process 
started at $x\in X_n$, 
so that,   
$P_x(\zeta^t=y)$ is the probability 
that the chain started at $x$ is in the configuration $y$ 
at time $t$ 
and 
$=P_x(\zeta^t\in Y)$ 
is the probability that a chain started at $x$ is in the set of 
configurations $Y\subset X_n$ at time $t$. 

In this paper, we will consider \gls{pca} defined as probabilistic
mixtures of \gls{eca}, namely, we shall assume 
\begin{equation}
\label{mod000}
p_i(y_i|x)
=
y_i\phi(x_{I_i})
+
(1-y_i)[1-\phi(x_{I_i})],
\end{equation}
where, given the positive integer $m$, 
the function
$\phi:Q^3\to[0,1]$ is defined as 
\begin{equation}
\label{mod010}
\phi=\sum_{r=1}^m \xi_r f_r,
\end{equation} 
with $\xi_r\in[0,1]$ such that $\sum_{r=1}^m\xi_r=1$
and $f_1,\dots,f_m$ are $m$ local rules.
It is important to note that the time evolution of a probabilistic 
mixtures of \gls{eca} can be described as follows: at time $t$ for each
cell $i\in\Lambda_n$ one chooses one of the rules $f_1,\dots,f_m$ with
probability $\xi_r$ and performs the updating based on the 
neighbourhood configuration at time $t-1$.
Indeed, with this algorithm the
probability to set the cell to $1$ a time $t$ is
the sum of the $\xi_r$ such that $f_i$, 
computed in the neighbourhood configuration at time $t-1$,
is one.

In the case $m=2$, which has been widely studied in the literature, 
see, e.g., \cite{CNS2021,Fautomata2017} and references therein, 
the probabilistic 
mixtures of \gls{eca} are called \emph{diploid} and equation 
\eqref{mod010}
is rewritten as 
\begin{equation}
\label{mod010bis}
\phi=(1-\lambda)f_1+\lambda f_2,
\end{equation} 
with $\lambda\in[0,1]$.

Finally, we remark that
the function $\phi:Q^3\to[0,1]$ can 
be interpreted as the probability to set the cell $i$ to $1$ 
given the neighbourhood configuration 
$x_{I_i}$ and, similarly, 
$1-\phi$ is the probability of selecting $0$.

\subsection{Order parameter}
\label{s:goal} 
\par\noindent
We are interested in studying the possibility of \gls{pca} to exhibit 
multiple stationary behaviours.
Given a measure $\mu$ on $Q$ which is stationary for 
the \gls{pca}, exploiting the translational invariance,
we will characterize it 
by the average occupation number of the cell at the origin 
\begin{displaymath}
\delta_\mu
=
\sum_{\newatop{x\in X_n:}{x_0=1}}
\mu(x),
\end{displaymath}
which will be called \emph{density}.

Sometimes, we will explore the stationary behavior 
using Monte Carlo simulations. In such a case we will 
start the \gls{pca} from an initial configuration in which 
cells are randomly and uniformly populated with zeros and ones. 
Then,
the density will be estimated by the large time behavior 
of the empirical average 
\begin{equation}
\label{e:density}
\delta_x(t)=\frac{1}{n}\sum_{i\in \Lambda_n } \zeta_i^t
.
\end{equation}

\subsection{Block approximation}
\label{s:block} 
\par\noindent
We propose a block approximation scheme of the stationary 
measure of any \gls{pca}. 
With the term \textit{block} we mean a sequence of contiguous cells.

Let us consider a block made of an odd number of contiguous 
cells, namely, the $(2m+1)$--{block} $B_{i,m}$ is the set of cells
$\{i-m,i-m+1,\dots,i,\dots,i+m-1,i+m\}$, with $i\in\Lambda_n$ and 
$1\le m\le(n-1)/2$; recall $\Lambda_n$ is an annulus and $n$ is 
an odd integer number.
Note that the neighborhood $I_i$ defined above is nothing 
but the block $B_{i,3}$.
Thus, given $z$ a configuration on $B_{i,m}$ we write 
\begin{align}
\label{eq:it010}
P_x(\zeta_{B_{i,m}}^t & =z) 
\notag\\
=&
\sum_{y\in X_n}P_x(\zeta_{B_{i,m}}^t=z|\zeta^{t-1}=y){P}_x( \zeta^{t-1}=y) 
\notag\\
=&
\sum_{y\in X_n} 
 \prod_{k=i-m}^{i+m} p_k(z_k|y)  P_x(\zeta^{t-1}=y) 
\notag\\
=&
\sum_{y\in Q^{B_{i,m+1}}} 
 \prod_{k=i-m}^{i+m} p_k(z_k|y)  
                     P_x(\zeta^{t-1}_{B_{i,m+1}}=y) 
.
\end{align}
Now, let $\varrho$ be the stationary 
distribution. We slightly abuse the notation by defining:
\begin{equation}
\label{eq:it012}
\varrho(z)
=
\sum_{y\in\Lambda_n:\, y_{B_{i,m+1}}=z} \varrho(y)
\end{equation}
for any $z\in Q^{B_{i,m}}$.
Thus, by setting $B_m=B_{0,m}$ and exploiting the translation invariance,
from \eqref{eq:it010}
we get
\begin{equation}
\label{eq:it014}
\varrho(z) 
=
\sum_{y\in Q^{B_{m+1}}} 
 \prod_{k\in B_m} p_k(z_k|y)  \varrho(y) 
\end{equation}
for any $z\in Q^{B_{m}}$.
We remark that the left-hand side of the equation above contains the $2^{2m+1}$ values of the stationary probability of the configurations on the $(2m+1)$--block $B_m$.
Instead, the right-hand side is written 
in terms of the $2^{2m+3}$ values of the stationary probability 
of the configurations on the $(2m+3)$--block $B_{m+1}$.

When considering a block approximation, the probabilities associated with the block $B_{m+1}$ are approximated by combinations of probabilities associated with smaller blocks.
If we denote by $P_\varrho$ these approximated block probabilities, the equations above can be closed to provide the approximated stationary maximal block probabilities as follows
\begin{equation}
\label{eq:it020}
\varrho(z) 
=
\sum_{y\in Q^{B_{m+1}}} 
 \prod_{k\in B_m} p_k(z_k|y)  P_\varrho(y) 
,
\end{equation}
where we recall
that $z\in Q^{B_m}$.

In the following we shall adopt the Bayes' rule for 
block approximation \cite{GVK87,GV87,FF15}, 
namely, 
\begin{equation}
\label{eq:it30}
P_\varrho(x_ix_{i+1})=\varrho(x_i)\varrho(x_{i+1}),
\end{equation}
and 
\begin{equation}
\label{eq:it40}
P_\varrho(x_i\dots x_{i+k})
=\frac{\varrho(x_{i}\dots x_{i+k-1})\varrho(x_{i+1}\dots x_{i+k})}
      {\varrho(x_{i+1}\dots x_{i+k-1})}
,
\end{equation}
for any $i$ and $k\ge2$. 
When the denominator vanishes, we assume $P_\varrho(x_i\dots x_{i+k})=0$.

In the applications that will be discussed in the following sections 
we shall use the approximation based on the maximal 
block of size $k$, for $k=3,5,\dots,13$, which will be called 
$k$--block approximation. 
It is useful to write explicitly the recursive equation that 
we find at least for the smaller values of $k$.
For the $3$-block approximation we have
\begin{align}
\label{eq:it050}
\varrho(z_{n-1}z_0z_{1}) 
&
=
\sum_{y\in Q^{B_2}} 
p_{n-1}(z_{n-1}|y)  
p_0(z_0|y)  
p_{1}(z_{1}|y)  
\vphantom{\bigg\{_\}}
\notag\\
&
\phantom{mm}
 \times
 \varrho(y_{n-2}y_{n-1}y_{0})\varrho(y_{n-1}y_{0}y_{1})
       \varrho(y_{0}y_{1}y_{2})
\notag\\
&
\phantom{mm}
 \times
 \frac{
       \mathbbm{1}_{\varrho(y_{n-1}y_{0})\varrho(y_{0}y_{1})>0}
       }
      {\varrho(y_{n-1}y_{0})\varrho(y_{0}y_{1})}
.
\end{align}
For the $5$-block approximation we have
\begin{align}
\label{eq:it060}
&
\varrho(z_{n-2}\dots z_{2}) 
\notag\\
&
\phantom{m}
=
\sum_{y\in Q^{B_3}} 
p_0(z_0|y)  
  \prod_{k=1}^2
   p_{n-k}(z_{n-k}|y)  
   p_{k}(z_{k}|y)  
\notag\\
&
\phantom{mm=}
 \times
 \varrho(y_{n-3}\dots y_{1})
       \varrho(y_{n-2}\dots y_{2})
       \varrho(y_{n-1}\dots y_{3})
\notag\\
&
\phantom{mm=}
\times
 \frac{
    \mathbbm{1}_{\varrho(y_{n-2}\dots y_{1})\varrho(y_{n-1}\dots y_{2})>0}
      }
      {\varrho(y_{n-2}\dots y_{1})
       \varrho(y_{n-1}\dots y_{2})}
.
\end{align}
For the $7$-block approximation we have
\begin{align}
&
\varrho(z_{n-3}\dots z_{3}) 
\notag\\
&
\phantom{mm}
=
\sum_{y\in Q^{B_4}} 
p_0(z_0|y)  
  \prod_{k=1}^3
   p_{n-k}(z_{n-k}|y)  
   p_{k}(z_{k}|y)  
\notag\\
&
\phantom{immmm}
 \times
 \varrho(y_{n-4}\dots y_{2})
       \varrho(y_{n-3}\dots y_{3})
       \varrho(y_{n-2}\dots y_{4})
\notag\\
&
\phantom{immmm}
 \times
 \frac{
    \mathbbm{1}_{\varrho(y_{n-3}\dots y_{2})\varrho(y_{n-2}\dots y_{3})>0}
      }
      {\varrho(y_{n-3}\dots y_{2})\varrho(y_{n-2}\dots y_{3})}
.
\end{align}
For the $9$--block approximation we have
\begin{align}
\label{eq:it090}
&
\varrho(z_{n-4}\dots z_{4}) 
\notag\\
&
\phantom{mm}
=
\sum_{y\in Q^{B_5}} 
p_0(z_0|y)  
  \prod_{k=1}^4
   p_{n-k}(z_{n-k}|y)  
   p_{k}(z_{k}|y)  
\notag\\
&
\phantom{immmm}
 \times
 \varrho(y_{n-5}\dots y_{3})
       \varrho(y_{n-4}\dots y_{2})
       \varrho(y_{n-3}\dots y_{3})
\notag\\
&
\phantom{immmm}
 \times
       \varrho(y_{n-2}\dots y_{4})
       \varrho(y_{n-3}\dots y_{5})
\notag\\
&
\phantom{immmm}
 \times
 \frac{
       \mathbbm{1}_{\varrho(y_{n-4}\dots y_{3})\varrho(y_{n-3}\dots y_{4})>0}
      }
      {\varrho(y_{n-5}\dots y_{3})
       \varrho(y_{n-3}\dots y_{3})}
.
\end{align}

Equations~\eqref{eq:it050}--\eqref{eq:it090} will be solved using 
the scheme proposed in \cite{K74} within the framework 
of the Cluster Variation Method and called the Natural Iteration Method, 
see, also, \cite{P05}.
More specifically, the initial $m$-block probabilities are chosen, and then, at each step of the iteration, the $(m-1)$-block probabilities are computed via symmetrized marginalization. 
Next, the equations~\eqref{eq:it050}--\eqref{eq:it090} are then employed to compute the new $m$-block probabilities. 
This algorithm is repeated till a fixed point 
is reached. In general we fix an \emph{a priori} 
precision of $10^{-7}$ for the $L^2$ norm of the $m$--block probabilities, 
but in some cases we shall discuss the dependence of the results on 
this parameter. 

\begin{figure}[H]
\includegraphics[width=0.22\textwidth]{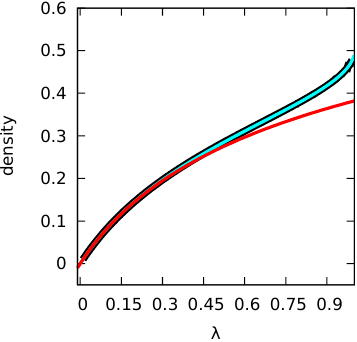}
\hskip 0.3 cm
\includegraphics[width=0.227\textwidth]{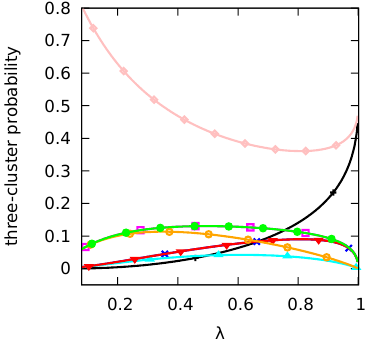}
\caption{\gls{ndeca} 17.
On the left, stationary density versus $\lambda$ computed
via mean field theory (red), $3$--block approximation (cyan),
and Monte Carlo simulation (black).
On the right,
$3$--block probability computed via the $3$--block 
approximation:
$\varrho(111)$ (black, cross),
$\varrho(110)$ (blue, times symbol),
$\varrho(101)$ (cyan, up triangle),
$\varrho(100)$ (magenta, open square),
$\varrho(011)$ (red, down triangle),
$\varrho(010)$ (orange, open disk),
$\varrho(001)$ (green, solid disk),
$\varrho(000)$ (pink, diamond).
}
\label{f:fig017}
\end{figure}

\section{Null rule diploid elementary cellular automata}
\label{s:ndeca} 
\par\noindent
In this section we focus our attention 
on mixtures of two elementary cellular automata in which 
one of the two mixed local rules is the null rule. 
More in the definition \eqref{mod010bis} of the function $\phi$ 
the map 
$f_1$ is the null rule (i.e., the local rule $W=0$) 
and $f_2$ is any other local rule.
As mentioned in the Introduction, 
following \cite{CNS2021},
we refer to these kinds of diploid as \gls{ndeca} 
and 
denote 
by \gls{ndeca} $W$ the \gls{ndeca} whose $f_2$ map is the local 
rule $W$.

We remark that the measure concentrated on the 
configuration $0$ is the unique
invariant measure for the finite--volume \gls{ndeca} in the 
cases in which the $f_2$ rule is even. Indeed, 
whatever is the configuration, 
cells with state one have a non--zero probability (at least $1-\lambda$)
to be changed to zero. 
Thus, if the zero rule is predominant, we expect a stationary 
measure essentially concentrated on the zero state. On the other hand, 
depending on the second local rule used in the mixtures, for 
$\lambda$ large enough, different behaviors can emerge.
Thus,
as discussed in \cite{CNS2021,Fautomata2017}, the main question 
is to understand if in the infinite-volume limit, namely, $n\to\infty$, 
a different stationary measure exists, with a positive value of the 
average cell occupation number. 

As mentioned in the Introduction,
this problem has been extensively studied in 
\cite{Fautomata2017} via numerical simulations and 
in \cite{CNS2021} via Mean Field approximation and rigorous one--site criteria. 
We shall approach here 
the problem by means of the block approximation for some particular \gls{ndeca} of interest. 
More specifically, we shall discuss those cases in which 
the Mean Field and the numerical predictions are not in agreement. 
All Monte Carlo results discussed in this section are obtained on the 
annulus $\Lambda_n$ with $n=10^5$ considering $2\times10^5$ full 
updates of the lattice starting from a random initial configuration
chosen with a Bernoulli distribution of parameter $1/2$.

In \cite{CNS2021} a detailed study of the Mean Field approximation has 
been provided and, in particular, it was proven that, within such an 
approximation, odd \gls{ndeca}s have a unique stationary state.
Such a result is consistent with the Monte Carlo study of 
\cite{Fautomata2017}.

The investigation of the particular case of the \gls{ndeca} 17 
via $3$ 
and $5$-block approximations confirmed the absence of a phase transition. 
Incidentally, such a \gls{ndeca} is known as the \emph{directed animals} 
\gls{pca} (see for instance Figure~7 in \cite{MM2014}) 
and it has been proven to have a unique invariant \emph{Markovian} measure.
The left panel of Figure~\ref{f:fig017} shows density as a function of  
$\lambda$. 
Results from the Mean Field match perfectly with those obtained using the 3-block approximation.
In the picture we have 
not reported graphs obtained with higher order block approximations, 
since they match perfectly with the $3$-block 
approximation prediction.

\begin{figure}[H]
\includegraphics[width=0.22\textwidth]{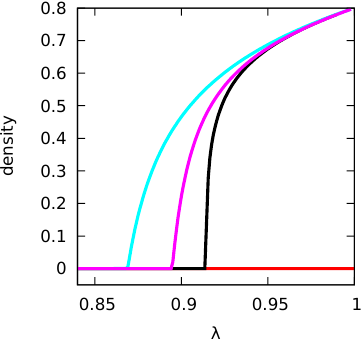}
\hskip 0.3 cm
\includegraphics[width=0.22\textwidth]{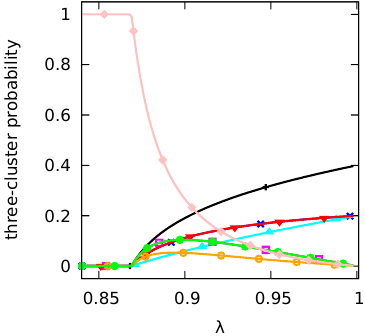}
\caption{As in Figure~\ref{f:fig017} for \gls{ndeca} 202. In the left 
panel the magenta line refers to the $5$--block approximation}
\label{f:fig202}
\end{figure}

When considering even \gls{ndeca}, Mean Field results are more interesting, 
even if less precise. 
As reported in \cite{CNS2021} within this approximation the non--uniqueness 
of the stationary measure 
is correctly predicted for all the models for which the simulation 
study of \cite{Fautomata2017} revealed this behavior, 
except for the \gls{ndeca} 202. 

As reported in Figure~\ref{f:fig202}, the existence of stationary 
measure not concentrated on $0$ 
is captured by the block approximations and the estimate of the value of the critical $\lambda$ gets closer to the Monte Carlo result as the order of the approximation is increased. 

In the right panel of Figure~\ref{f:fig202}, we report the $3$-block probability computed via the $3$-block approximation.
It is interesting to remark that at large values of $\lambda$ the $3$-block probability does not concentrate on the $111$ configuration. 
This explains why Mean Field is not able to capture the 
transition. 

\begin{figure}[H]
\includegraphics[width=0.22\textwidth]{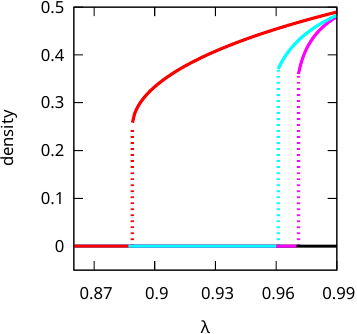}
\hskip 0.3 cm
\includegraphics[width=0.22\textwidth]{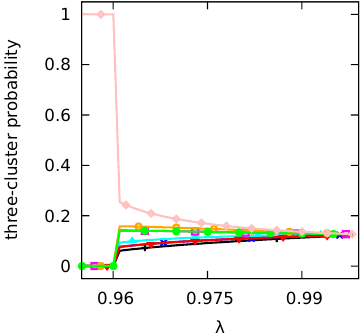}
\caption{As in Figure~\ref{f:fig202} for \gls{ndeca} 120.}
\label{f:fig120}
\end{figure}

On the other hand, there are also cases where the Mean Field 
theory predicts the existence of a transition that is not observed 
in Monte Carlo simulations, such as in the case of \gls{ndeca} 120.
As reported in Figure~\ref{f:fig120}, the Mean Field predicts a discontinuous 
transition, which is confirmed by the block approximations, 
but only for values of $\lambda$ very close to unity.
This evidence, together with the fact that all $3$-block 
probabilities converge to the same value for $\lambda$ 
large (see the graphs on the right panel of the same figure), 
strongly suggests that this is an artifact of the block approximations. 

\begin{figure}[H]
\includegraphics[width=0.22\textwidth]{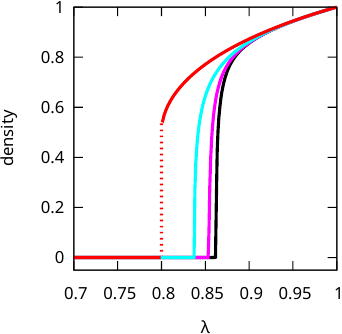}
\hskip 0.3 cm
\includegraphics[width=0.228\textwidth]{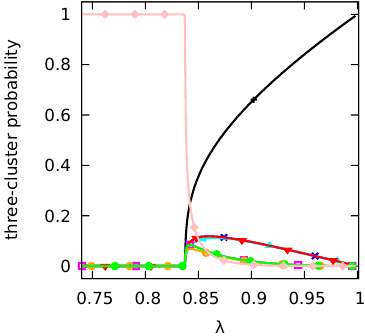}
\caption{As in Figure~\ref{f:fig202} for \gls{ndeca} 234.}
\label{f:fig234}
\end{figure}

Mean Field approximation finds a discontinuous transition also
for \gls{ndeca} 234, for which Monte Carlo simulation predicts the existence of a continuous transition. 
As illustrated in Figure~\ref{f:fig234}, the block approximations recover a 
continuous transition and, as already remarked in a similar case, the estimate of the critical value of $\lambda$ gets closer and closer to the Monte Carlo prediction when the order of the block approximation is increased. 
It is also worth noting that the Mean Field and the block 
approximation estimate of the density matches the Monte Carlo prediction perfectly for large $\lambda$: this is the case when the block probabilities concentrate on the configuration $111$, as shown in the right panel of the figure. 

\begin{figure}[H]
\includegraphics[width=0.22\textwidth]{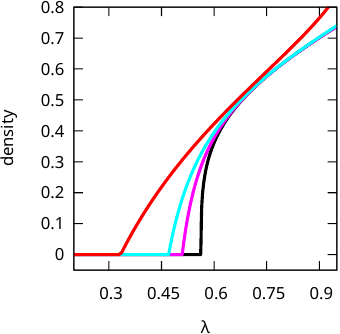}
\hskip 0.3 cm
\includegraphics[width=0.23\textwidth]{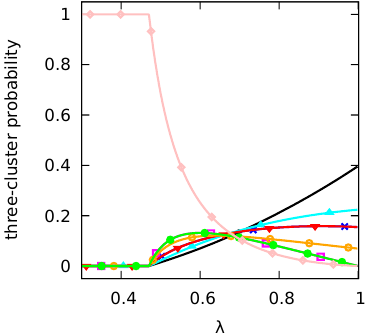}
\caption{As in Figure~\ref{f:fig202} for \gls{ndeca} 222.}
\label{f:fig222}
\end{figure}

As reported in \cite{CNS2021}, Mean Field also predicts the existence of 
the transition for the \gls{ndeca} 222, a phenomenon not investigated in \cite{Fautomata2017}.
This is confirmed by the results of Figure~\ref{f:fig222}. 
Interestingly, we remark that the density profile at large $\lambda$ is not well reproduced through the Mean Field, whereas a perfect match is recovered when block approximations are considered.
As noticed, this is imputed to the fact that when $\lambda$ is close 
to one the $3$-block probabilities do not concentrate on $111$; see the right panel of Figure ~\ref{f:fig222}.
Such a  situation is not easily captured by Mean Field. 
%
Indeed, the main limit of the Mean Field approximation is due to the fact that the equation providing the stationary density depends only on the total number of configurations with a fixed number of ones that associate 1 to the configuration in the neighborhood under the map $f$. 

\begin{figure}[H]
\includegraphics[width=0.22\textwidth]{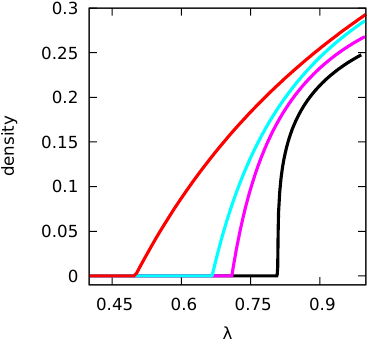}
\hskip 0.3 cm
\includegraphics[width=0.22\textwidth]{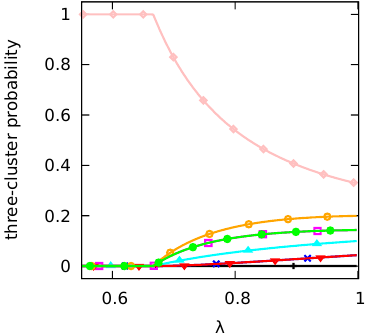}
\caption{As in Figure~\ref{f:fig202} for \gls{ndeca} 18.}
\label{f:fig018}
\end{figure}

We now pass to discuss other useful examples to illustrate how the 
block approximation works.
The \gls{eca} 18 is a chaotic \gls{ca} belonging to Wolfram's class $W3$;
it is also called \emph{diffusive rule}.
The main feature of \gls{eca} 18 is that at time $t$ it 
updates a central site with a one only if at time $t-1$ there is a one either on its left or on its right.  
Thus, it is an example of a symmetric rule.
Figure~\ref{f:fig018} shows that also this \gls{ndeca} does not concentrate, for $\lambda$ close to unity, the 3-block probabilities on the outcome $111$.
As a consequence, the Mean Field prediction is not particularly satisfactory, whereas larger block approximations are able to predict a density profile that gets closer and closer to the Monte Carlo result. 
It is worth noting that this diploid is particularly challenging for the block approximations, as shown by the graphs in the left panel of the figure, since the $111$ block configuration does not contribute to the stationary density while all the other $3$--block configurations have 
not zero stationary probability. 

\begin{figure}[H]
\includegraphics[width=0.22\textwidth]{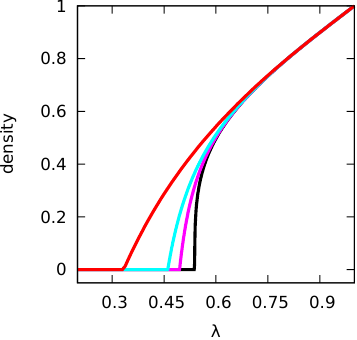}
\hskip 0.3 cm
\includegraphics[width=0.226\textwidth]{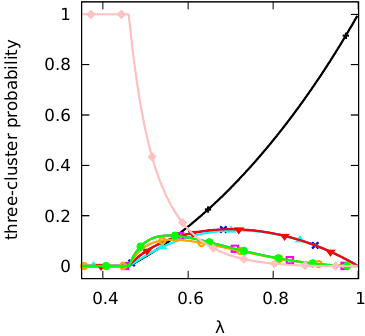}
\caption{As in Figure~\ref{f:fig202} for \gls{ndeca} 254.}
\label{f:fig254}
\end{figure}

\begin{figure}[H]
\includegraphics[width=0.22\textwidth]{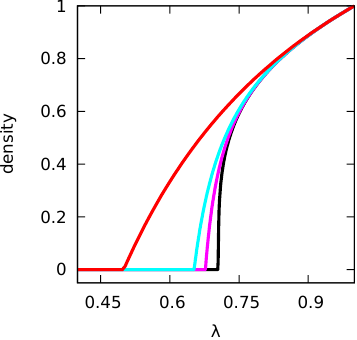}
\hskip 0.3 cm
\includegraphics[width=0.226\textwidth]{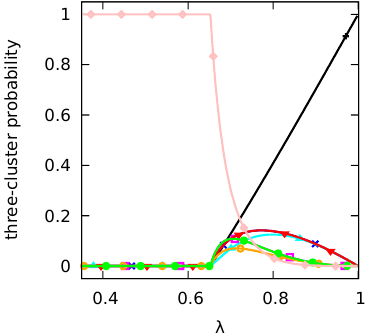}
\caption{As in Figure~\ref{f:fig202} for \gls{ndeca} 238.}
\label{f:fig238}
\end{figure}

The \gls{ndeca} 254 is based on the simple rule 254, which has the configuration with all ones as fixed point (Wolfram's class $W1$).
For this diploid, see Figure~\ref{f:fig254}.
The Mean Field prediction and the Monte Carlo results compare very well far from the critical point.
Once again, this is due to the fact that for a $\lambda$ close to unity the block probabilities concentrate on $111$. 
This diploid is well known in the literature, see, e.g.,
\cite{BMM2013,T2004}, where it is called \emph{percolation} \gls{pca}.  
In \cite[Example~2.4]{BMM2013} the existence of the 
transition is proven rigorously. 
The exact value of the critical $\lambda$ is not known, but it 
is proven that it belongs to the interval
$[1/3, 53/54]$, see \cite{T2004}. 
In \cite{taggi} the lower bound $0.505$ is given.  
Mean Field, $3$--block, and $5$--block approximations, respectively, predict $1/3$, $0.461\pm0.001$, and $0.496\pm0.001$.
For the block approximations, we take as the operative definition of critical $\lambda$ the value such that the density is larger than $0.01$.

As a last example, we consider the \gls{ndeca} 238, whose behavior, very similar to that of \gls{ndeca} 254, is reported in Figure~\ref{f:fig238}.
This \gls{ndeca} is called the \emph{Stavskaya} model and it is a particular case of the \emph{percolation} \gls{pca}. 
For the critical value of $\lambda$, the lower bound is $0.677$ (see \cite{MScmp1991}). 
Mean Field, $3$--block, and $5$--block approximations respectively predict $0.5$, $0.654\pm0.001$, and $0.678\pm0.001$.

\section{A diploid with two absorbing states}
\label{s:deca_182_200} 
\par\noindent
In this section, we consider the diploid 
obtained as a 
mixtures of the local rules $W=200$ and $W=182$ \cite{Me2011a}, see 
Table~\ref{tab:rules_182_200}.
More precisely, the model is defined 
by the equations \eqref{mod000a}, \eqref{mod000}, and 
\eqref{mod010bis}
where
$f_1$ is the local rule 200 and 
$f_2$ the local rule 182.
The dynamics becomes deterministic as $\lambda$ assumes 
the values  0 and 1 and, in particular, it reduces 
to the \gls{eca} 200 and 182, respectively.

As mentioned above, we use our algorithmic implementation of the 
block approximation to study the phase diagram of the model and 
we show that, provided the size of the maximal block is sufficiently
large, the main features of the Monte Carlo results are fully recovered.

\begin{table}[H]
    \centering
    \begin{tabular}{|l|cccccccc|}
    \hline
         & 111 & 110 & 101 & 100 & 011 & 010 & 001 & 000\\
        \hline
        \textbf{200} & 1 & 1 & 0 & 0 & 1 & 0 & 0 & 0\\
        \textbf{182} & 1 & 0 & 1 & 1 & 0 & 1 & 1 & 0 \\
        \hline
    \end{tabular}
    \caption{Table with deterministic updating dynamics for rules 182 and 200.
    The first column gives the rule, while the others give the updating of the central site given its neighbors which are reported in the first row.}
    \label{tab:rules_182_200}
\end{table}

As a first remark, we note that this diploid has two absorbing states
halting any further configuration development. These are the 
configurations 
$0$ and $1$, as it clearly follows from the fact that both the 
local rules 200 and 182 associate $1$ to the neighbourhood
configuration $111$ and $0$ to the neighborhood configuration
$000$, see Table~\ref{tab:rules_182_200}.
Moreover, for $\lambda= 0$, 
the diploid reduces to the \gls{eca} 200 which is generally considered 
uninteresting in cellular automata studies because most 
initial configurations either die out quickly or stabilize into 
simple, unchanging patterns. As remarked in \cite{Me2011a}, see 
also \cite[Table~6 on page 540]{W94}, if the automaton is started 
with an initial configuration in which zeros and ones are chosen 
with probability $1/2$, then the dynamics reaches a stationary 
density value equal to $3/8$.
On the contrary, 
for $\lambda= 1$, the diploid is nothing but \gls{eca} 182, which 
results in a stationary density, independently on the 
starting configuration, equal to $3/4$ 
\cite[Figure~12 and page~23]{W94} and \cite{Grassberger:1982}.
We recall that this \gls{eca}, 
although it can produce interesting behaviors, it generally 
leads to predictable and orderly patterns.
Another peculiar feature of the diploid object of our study
is observed 
at $\lambda=1/2$, where the updating dynamics of any 
site becomes no longer correlated with respect to its neighbors, 
excepted for the $111$ ad $000$ neighborhood states. 
Indeed, in all other cases the function $\phi$ 
takes values $1/2$. In other words, 
in the absence of long sequences of zeros or ones, 
sites randomly flip with probability $1/2$, so that
the stationary density is expected to fluctuate around $1/2$
\cite{Me2011a}.

As repeatedly recalled, 
the diploid 182--200 has been studied in \cite{Me2011a}, 
where Mean Field, Monte Carlo simulations, and 
small block approximations (up to $4$)
were used to investigate the phase diagram. 
Here we exploit our algorithmic version of the 
block approximation to push further the analysis using 
maximal block size up to $13$.

We summarize briefly the results in \cite{Me2011a}, 
Monte Carlo simulations revealed that 
there is a region in which the stationary state 
is given by the $0$ absorbing state, this happens for 
$\lambda$ approximatively in the interval $(0.03,0.48)$. 
On the other hand, no region in which the $1$ absorbing state provides the stationary behaviour is found. 
For $\lambda<0.03$ and $\lambda>0.48$ two active regions 
are detected, meaning that the stationary 
density is positive. 
For $\lambda>0.48$ the density increases up to the stationary 
solution $3/4$ correctly found at $\lambda=1$.
The density profile shows a local maximum for $\lambda$ close to 
$0.58$, an inflection point around $0.62$, and a local minimum 
around $0.75$.
Mean Field predicts an active state 
with density  $(3\lambda-1)/(4\lambda-1)$ as long as 
$\lambda\ge1/3$, which means density at $\lambda=1$ equal 
to $2/3$ (instead of $3/4$), 
transition point at $1/3$ (instead of $0.48$), and 
no active state at small values of $\lambda$.
On the other hand, 
it correctly predicts the value $1/2$ for the density
at $\lambda= \frac{1}{2}$.
Finally, it was also remarked that 
considering block approximation with maximal block size 
up to $2,3,4$ provides increasingly better estimates 
for the critical point and the value of the density at $\lambda=1$. 

\begin{figure}[H]
    \centering
    \includegraphics[width=0.45\textwidth]{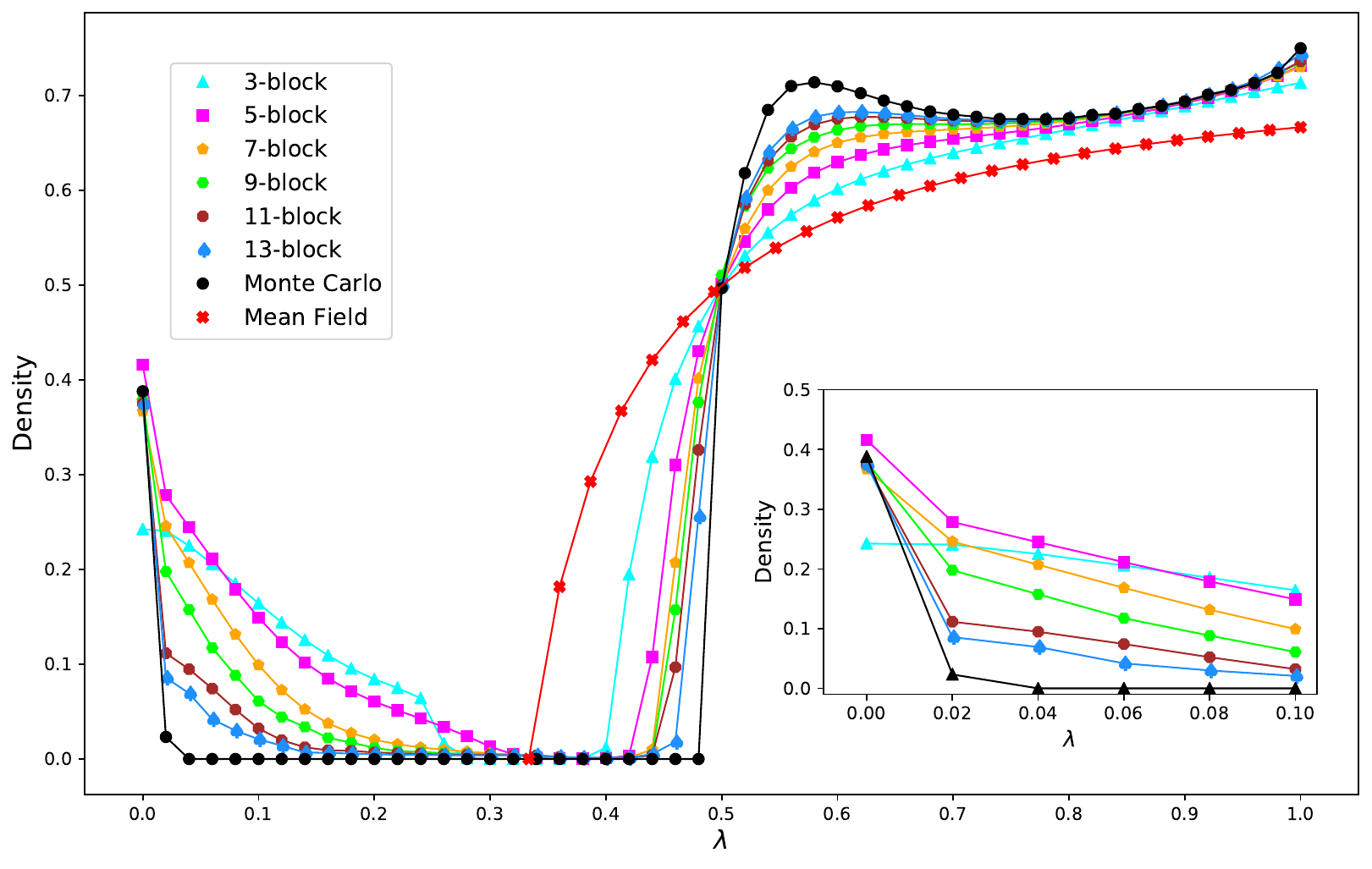}
    \caption{Density profile for the diploid 200-182 as 
     a function of $\lambda$.
    Colour code: 
    Monte Carlo is shown in black,
    3-block approximation in cyan, 
    5-block in magenta, 
    7-block in orange,  
    9-block in green,  
    11-block in brown, 
    and 13-block in blue.
    For each block approximation, the same randomly initialized density was used; for the Monte Carlo simulations, sites were randomly initialized with density 1/2.
The inset is a magnification of the plot within the range
$[0,0.1]$.
}
    \label{f:182-200}
\end{figure}

In this paper, we study the diploid $200$--$182$ using the block approximation
with maximal block size up to $13$ and in 
Figure~\ref{f:182-200} we report, as a function of $\lambda$, 
the density computed within this approximation together with 
the results of the Monte Carlo simulation.
The block approximation recursive equation has been solved 
iteratively starting from an initial condition in which the 
maximal block density function was chosen randomly.

To ensure the reproducibility of results, we
used the same random initial condition across all simulations. 
Specifically, the initial block density was generated by drawing 
uniformly distributed positive values and then normalizing them. 
We opted for this
method to prevent any bias from a specific initial configuration and to
simulate a neutral starting point that fairly represents a broad range of
possible initial states. 
In fact, this strategy was chosen to mitigate the
risk of the algorithm converging on a distribution with most of its mass
on either the zero or the one configuration, which would mirror the 
absorbing states of 0 and 1 in the diploid. As a result, such an approach 
allowed
us to explore the stationary distribution more effectively.

Overall, higher-order block approximations provide a more accurate 
description of the Monte Carlo results 
compared to the low-order. 
But let us discuss our findings in detail. 

For $\lambda$ small, unlike Mean Field, the block 
approximation is able to reproduce the active state found in the Monte Carlo and the steep descent to zero is more and more accurately described when the size of the approximation is increased. 
We stress that the Mean Field approach was not able to describe the non-trivial behaviour at small values of $\lambda$, while the block approximation method proves to be more effective and reliable in the inspection of this domain.
We add that the block approximation equations
return the solution concentrated on the zero configuration if 
the solving iterations are initialized with a block matrix 
close to the Kronecker delta on the zero block configuration. 
Likewise, starting the iteration with an
opposite initialization close to the Kronecker delta on the 1 block 
configuration, results in a solution concentrated on the 1 block 
configuration itself.

The existence of the transition at $0.48$ is confirmed by the 
block approximation and the estimate of the transition 
point is more and more accurate 
as the order of the approximation is increased.
Indeed, for the transition point 
within the 3-block approximation, we find $0.4$, 
within the 5-block approximation $0.42$, 
within both the 7-block and 9-block approximations $0.44$, and
within the 11-block and 13-block approximation $0.46$.

At the notable point $\lambda= 1/2$, as for the Mean Field, 
all the block approximation-based densities achieve 
values close to $1/2$, confirming the Monte Carlo and 
the symmetry argument prediction.

The region $\lambda>1/2$ is particularly interesting, indeed, the competition between the two local maps defining the diploid produces a not monotonic graph of the stationary density. 
This behaviour is not at all captured by the Mean Field approximation 
and is retrieved only by the block approximation with a large 
maximal block size.

More precisely, all block approximations up to maximal block size 9 show
monotonic curves, but a concavity change can be observed at $\lambda = 0.68$ and
$\lambda = 0.70$ in the 7-block and 9-block cases, respectively. 
For the 11-block, we observe a local maximum at $\lambda = 0.62$, an inflexion point at $\lambda = 0.67$, and a local minimum at $\lambda= 0.74$.

As said above, all our results have been found solving 
the iterative equations with precision $10^{-7}$.
For completeness, we investigate 
the sensitivity of the block approximation method on 
the choice of this technical parameter.
Unlike \gls{ndeca}, where the null rule significantly contrasts 
with the other rule as $\lambda$ approaches 0, 
the diploid 200--182 exhibits, for small $\lambda$, an active region.
More specifically, \gls{eca} 200 is expected to be 
more prevalent than \gls{eca} 182 below the transition value of $\lambda$.
Consequently, the dynamics 
approaches some stationary state made of bulks of ones and zeros.
In such a scenario, the presence of \gls{eca} 182 
allows for the flipping of a single site within one bulk, causing the collapse of that block to zero when \gls{eca} 200 is applied subsequently.
As a result, the dynamics fall into the null absorbing state as the contrast between the two rules is mild (say, $\lambda$ around $0.3$), while non-null stationary measures become more prevalent as the contrast becomes weaker 
(say, $\lambda$ close to zero).
Heuristically, we can say that this contrast may prematurely halt the iterative search of block approximation algorithms. 
This occurs because \gls{eca} 182 may not be selected frequently enough to guide the search towards the Monte Carlo solution.

\begin{figure}[H]
    \centering
\includegraphics[width=0.45\textwidth]{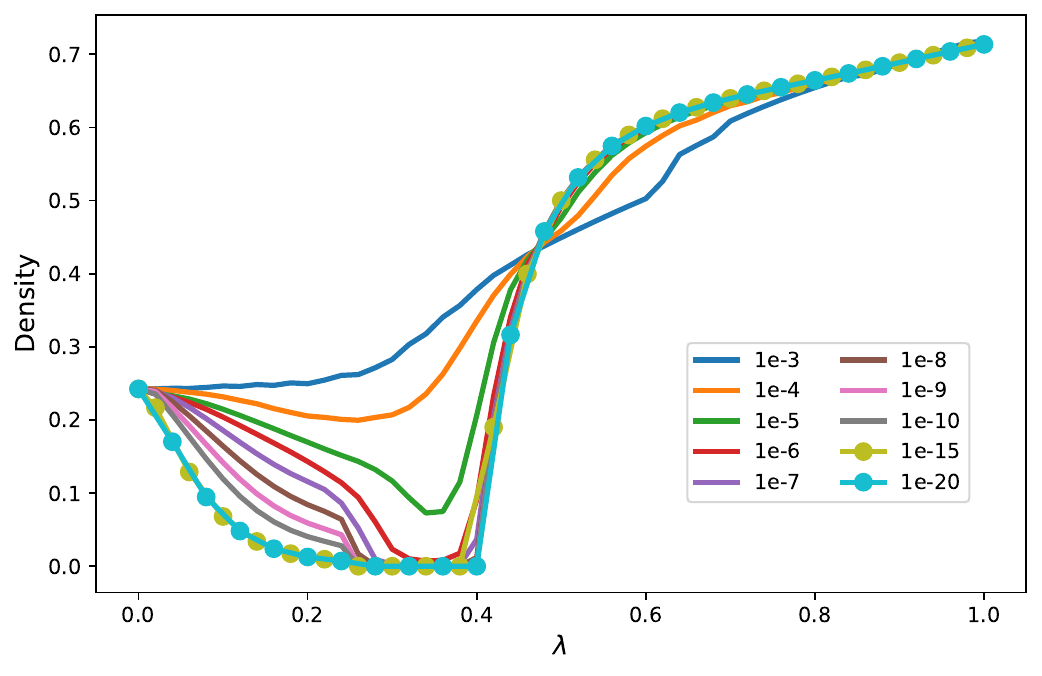}
    \caption{Sensitivity analysis of the 3-block approximation with 
respect to the precision of the algorithm. Colored lines show approximations at different values of of precision}
    \label{f:sensitiviy_analysis_block_3}
\end{figure}

In figure \ref{f:sensitiviy_analysis_block_3} the 3-block approximations are shown at various levels of precisions.
Specifically, we explored the level of precisions from a maximum of $10^{-3}$ up to a minimum of $10^{-20}.$
As for the analysis of Figure~\ref{f:182-200}, we used 
the same initial conditions for each run to give this analysis more reproducibility. 
As expected, the accuracy of the block approximation improves as the algorithm's precision decreases. 
Such an improvement is particularly noticeable away from $\lambda = 0$, where the transition from the equilibrium state of \gls{eca} 200 to the null absorbing state is better captured when higher precision levels are selected for the block approximation.
In the region $\lambda>1/2$ the precision $10^{-4}$ is already sufficient to get very precise solutions of the iterative equations, 
on the other hand in the region close to $0$ the precision should 
be pushed to $10^{-15}$. 
Thus, we have to conclude that in this region the curves discussed in Figure~\ref{f:182-200} are only qualitatively reliable.

\section{An example of one--dimensional totalistic \gls{pca}}
\label{s:1d_bbr}
\par\noindent
In this section, we consider 
the model which is the main object of study in \cite{Ba2000}.
It is 
an example of 
totalistic \gls{pca}, namely, a \gls{pca} 
in which the one--site updating probability $p_i(\cdot|x)$ depends 
only on the total of the values of the cells in the neighborhood
$I_i$, namely, on $\sigma_i(x)=x_{i-1}+x_i+x_{i+1}$.
From our point of view, the interest of the model is twofold:
i) it shares with the diploid studied in Section~\ref{s:deca_182_200}
the property that it admits two absorbing states, i.e., $0$ and $1$;
ii) it is an example of probabilistic mixtures of four \gls{eca}.
In this complicated setup we 
show the ability of the block approximation to provide a very precise 
description of a highly non--trivial phase diagram.

Following \cite{Ba2000} and recalling the notation introduced 
in Section~\ref{s:mixture} the totalistic \gls{pca} is defined 
by saying that the cell variable at time $t$ is set 
to $0$ if $\sigma_i(\zeta^{t-1})=0$, 
to $1$ if $\sigma_i(\zeta^{t-1})=3$, 
and it is chosen by performing a Bernoulli trial with 
parameter $q_s$ if
$\sigma_i(\zeta^{t-1})=s$, 
with $s=1,2$, and $q_1,q_2\in[0,1]$.
This is equivalent to say that the one--site updating probability 
are the following
\begin{align}
\label{ba000}
&
p_i(1|000)=0,\notag\\
&
p_i(1|001)=p_i(1|010)=p_i(1|100)=q_1,\notag\\
&
p_i(1|011)=p_i(1|101)=p_i(1|110)=q_2,\notag\\
&
p_i(1|111)=1,
\end{align}
for any $i\in\Lambda_n$, where we misused the notation introduced in 
\eqref{mod000} by 
reporting on the right of the bar only the neighborhood configuration.

\begin{table}[H]
    \centering
    \begin{tabular}{|l|cccccccc|}
    \hline
         & 111 & 110 & 101 & 100 & 011 & 010 & 001 & 000\\
        \hline
        \textbf{254} & 1 & 1 & 1 & 1 & 1 & 1 & 1 & 0\\
        \textbf{232} & 1 & 1 & 1 & 0 & 1 & 0 & 0 & 0\\
        \textbf{150} & 1 & 0 & 0 & 1 & 0 & 1 & 1 & 0\\
        \textbf{128} & 1 & 0 & 0 & 0 & 0 & 0 & 0 & 0 \\
        \hline
    \end{tabular}
    \caption{Local rules used in the definition of the model 
in Section~\ref{s:1d_bbr}.
    The first column reports the local rule decimal code, 
 while the others report the updating of the central site given its neighbors which are listed in the first row.}
    \label{tab:rules_bbr}
\end{table}

It is very interesting to remark that this model can be rewritten 
as the probabilistic mixture of \gls{eca} \eqref{mod000}--\eqref{mod010}
of the four local rules $254$, $232$, $150$, $128$, respectively,
as the maps $f_1$, $f_2$, $f_3$, $f_4$ in \eqref{mod010} with 
the four reals $\xi_1,\dots,\xi_4\in[0,1]$ satisfying 
the equations
\begin{equation}
\label{ba010}
\xi_1+\xi_2=q_2,\,
\xi_1+\xi_3=q_1,\,
\xi_1+\xi_2+\xi_3+\xi_4=1,
\end{equation}
and 
the four local rules considered above 
detailed in Table~\ref{tab:rules_182_200}.
Indeed, \eqref{ba000} are recovered noting that 
from \eqref{mod000} and \eqref{mod010} one has 
$p_i(1|x)
=
\phi(x)
=
\sum_{r=1}^4
\xi_r f_r$, 
and thus 
\begin{align}
&
p_i(1|000)=0,\notag\\
&
p_i(1|001)=p_i(1|010)=p_i(1|100)=\xi_1+\xi_3,\notag\\
&
p_i(1|011)=p_i(1|101)=p_i(1|110)=\xi_1+\xi_2,\notag\\
&
p_i(1|111)=\xi_1+\xi_2+\xi_3+\xi_4.\notag
\end{align}
Finally, we remark that the system of equations \eqref{ba010} is compatible 
in $[0,1]$ and a solution can be written as follows: 
\begin{align}
\label{ba020}
&\xi_1=q_1,\xi_2=q_2-q_1,\xi_3=0,\xi_4=1-q_2
&\textrm{if }q_1<q_2,
\notag\\
&\xi_1=q_1,\xi_2=\xi_3=0,\xi_4=1-q_1
&\textrm{if }q_1=q_2,
\notag\\
&\xi_1=q_2,\xi_2=0,\xi_3=q_1-q_2,\xi_4=1-q_1
&\textrm{if }q_1>q_2.
\notag
\end{align}

The Mean Field computation in \cite{Ba2000}
provides for the stationary density the solutions $0$, $1$, and 
$\delta(q_1,q_2)=(3q_1-1)/(1+3q_1-3q_2)$. The last one is meaningful, 
namely, it takes values in $[0,1]$, only for  
$q_1\ge1/3$ and $q_2 \le 2/3$.
Since $\delta(1/3,q_2)=0$ and $\delta(q_1,2/3)=1$, 
we can say that
the Mean Field approximation  
predicts the following phase diagram in the 
square $[0,1]\times[0,1]$:
the absorbing states 0 and 1 are respectively the stationary states 
in the region $q_1<1/3$ and $q_2<2/3$
and 
in the region $q_1>1/3$ and $q_2>2/3$.
On the other hand, in 
the domain $1/3\le q_1$ and $2/3\ge q_2$ 
the stationary density is provided by the continuous function 
$\delta(q_1,q_2)$.
Nothing can be said in the squared region $q_1<1/3$ and $q_2>2/3$.

\begin{figure}[H]
    \centering
    \includegraphics[width=0.45\textwidth]{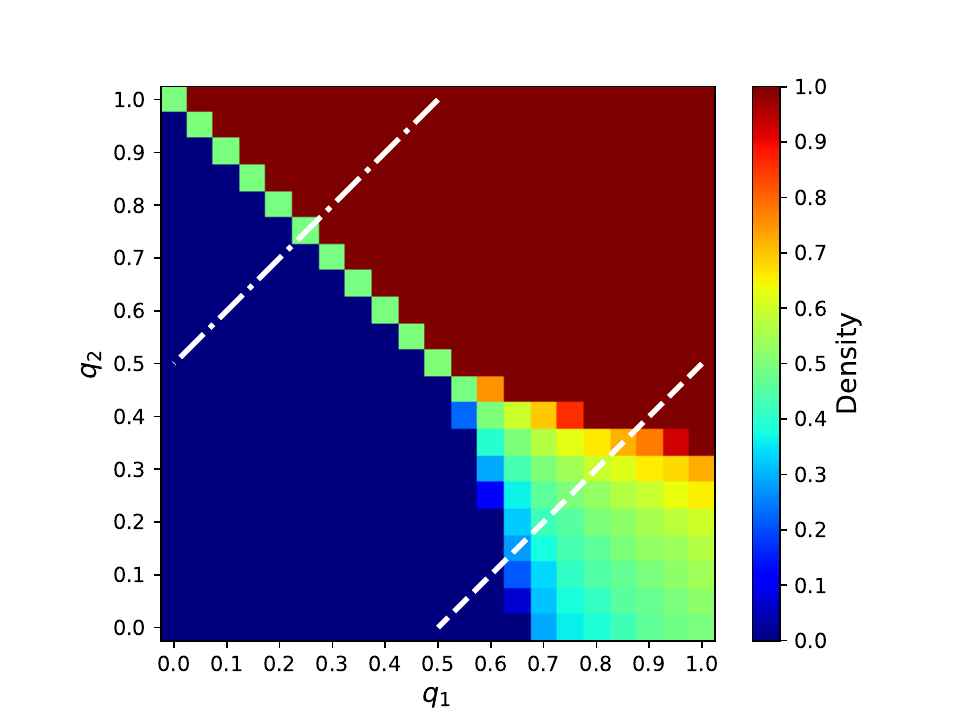}
    \caption{Monte Carlo simulation of the one-dimensional totalistic 
     \gls{pca}. 
    Parameters $q_1$ and $q_2$ are shown along the horizontal and vertical 
    axes, respectively.
    The heatmap shows the density for a given couple 
    of parameters $(q_1, q_2)$ on a grid of $21\times21$ points 
    in $[0,1]\times[0,1]$ with spacing $0.05$.
    The dashed-dotted and the dashed lines represent respectively the line of equations $q_2 = q_1 + 1/2$ and  $q_2 = q_1 - 1/2$.}
    \label{fig:BBR_MC}
\end{figure}

The Monte Carlo density prediction, which is in perfect agreement with that
reported in \cite[Figure~3]{Ba2000}, is 
shown in Figure~\ref{fig:BBR_MC}. 
Simulations were accomplished on a 10000-length array 
throughout 10000 iterations.
Monte Carlo simulations reveal a scenario that confirms 
only partially the Mean Field predictions.
In particular, close to the point $q_1=0$ and $q_2=1$, 
we observe a discontinuous transition between the $0$ and $1$ states, 
indeed the density has an abrupt jump from $0$ to $1$. 
More precisely, 
this discontinuous transition is observed through the line 
$q_1 + q_2 = 1$ for $q_1\in(0,1/2)$.

It is interesting to remark that, exactly on the line 
$q_1 + q_2 = 1$ with $q_1\in(0,1)$, the observed value for the stationary 
density is $1/2$. This fact is not surprising since for this choice 
of the parameters 
the totalistic model reduces to the diploid studied in 
Section~\ref{s:deca_182_200}
at $\lambda=1/2$. 

The discontinuous transition dies out at $q_1=1/2$ where the 
discontinuous transition line 
bifurcates and gives rise to two curves, symmetric with respect
to the line $q_1+q_2=1$, 
where a continuous transition between, respectively, the 
$0$ and $1$ absorbing states and a continuous phase with positive density 
are observed. 

In order to study this interesting phase diagram with the 
block approximation   to confirm the Monte Carlo results 
both in the discontinuous and continuous transition regions, we have 
we explored the model along the 
lines $q_2 -q_1=1/2$ and  $q_2-q_1=-1/2$ (see the dotted 
dashed and dotted lines in Figure~\ref{fig:BBR_MC}).

\begin{figure}[H]
    \centering
    \includegraphics[width=0.45\textwidth]{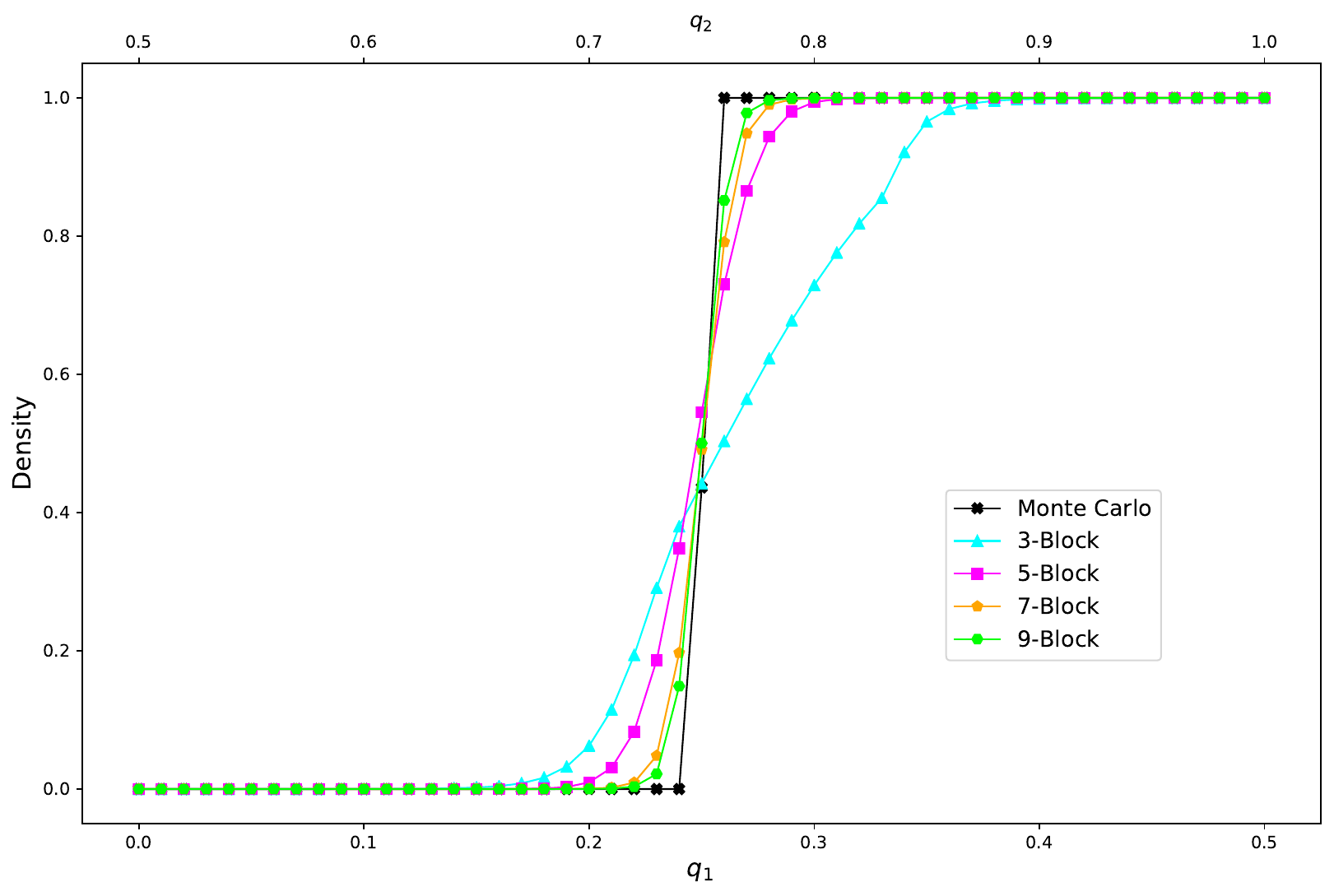}
    \caption{Stationary density along the line $q_2 -q_1=1/2$.
    In cyan the 3-block, 
    in magenta the 5-block, 
    in orange the 7-block, 
    and in green the 9-block approximation.
    In black the Monte Carlo simulation.
    On top and on the bottom axes, the values of $q_2$ and $q_1$ are 
    respectively reported.
    }
    \label{fig:BBR_MC_line_1}
\end{figure}

In Figure~\ref{fig:BBR_MC_line_1}, Monte Carlo and block approximation 
estimates for the stationary density along the line $q_2 -q_1=1/2$ are 
shown.
The model exhibits a discontinuous transition, 
from the quiescent zero density state to the active density one state
at $q_1= 1/4$, or equivalently, $q_2=3/4$.
This behaviour is clearly captured by the Monte Carlo simulations and it is 
clearly confirmed by a careful 
analysis of the block approximation results.
The 3-block approximation predicts a very smooth change from the 
quiescent to the active state and the transition point can be 
estimated at $q_1 = 0.18$ (or $q_2 = 0.68$).
A more sudden change is observed with the 5-block approximation and the 
transition point estimate becomes $q_1 = 0.2$ (or $q_2 = 0.7$).
The density curves computed with the $7$--block and $9$-block are 
almost perfectly overlapped: the description of the sharp transition 
is more accurate and the transition point is estimated at 
$q_1 = 0.22$ (or $q_2 = 0.72$).

\begin{figure}[H]
    \centering
    \includegraphics[width=0.45\textwidth]{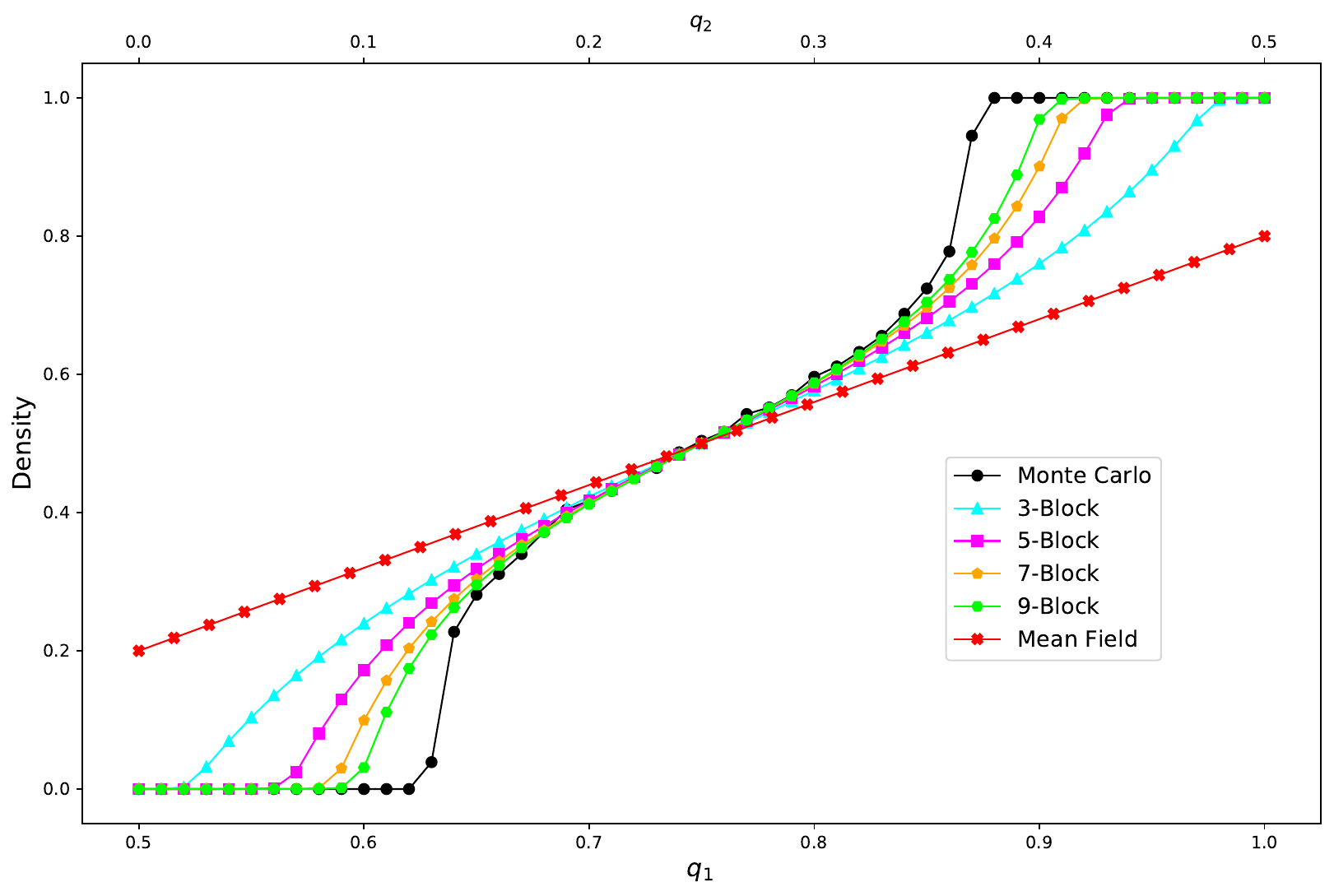}
    \caption{Stationary density along the line $q_2 -q_1=-1/2$.
    In red the Mean Field computation.
    In cyan the 3-block, 
    in magenta the 5-block, 
    in orange the 7-block, 
    and in green the 9-block approximation.
    In black the Monte Carlo simulation.
    On top and on the bottom axes, the values of $q_2$ and $q_1$ are, respectively, reported.
    }
    \label{fig:BBR_MC_line_2}
\end{figure}

In Figure~\ref{fig:BBR_MC_line_2}, Mean Field, 
Monte Carlo, and block approximation 
estimates for the stationary density along the line $q_2 -q_1=-1/2$ are 
reported.
The Monte Carlo estimate of the stationary density 
reveals the presence of two continuous transitions between the 
two absorbing states, namely,  $0$ and $1$, and a continuous phase 
with positive density.
The two transition points are close to 
$q_1=0.63$ (or $q_2=0.13$) and 
$q_1=0.87$ (or $q_2=0.37$). 
In between, the stationary density varies smoothly. 
The Mean Field approximation is not able to capture the two 
continuous transitions, although the agreement with the Monte Carlo 
results is good close to $q_2=3/4$ where the line 
$q_2 -q_1=-1/2$ intersect the $1/2$ density line 
$q_1+q_2=1$.

On the other hand, block approximations provide a precise description of the two continuous transitions.
Indeed, the agreement between Monte Carlo and block 
approximation prediction is perfect in the centre part 
of the graph, namely, around $q_1= 0.75$ (or $q_2 = 0.25$). 
On the other hand, when the transition point is approached, 
the block approximation curves depart from the Monte Carlo 
density prediction, but the estimate of the transition 
points get better and better when the size of the maximal block is increased. 
More precisely, 
for what concerns the transition point involving the quiescent $0$ state, 
the 
$3$--block, 
$5$--block, 
$7$--block, 
$9$--block,
respectively, predicts
$q_1=0.52$, $q_1=0.56$, $q_1=0.58$, and $q_1=0.59$.
On the other hand, for the transition point involving the active $1$ state, 
the analogous sequence is 
$q_1=0.97$, $q_1=0.94$, $q_1=0.92$, and $q_1=0.91$.

\begin{figure}[H]
\includegraphics[width=0.22\textwidth]{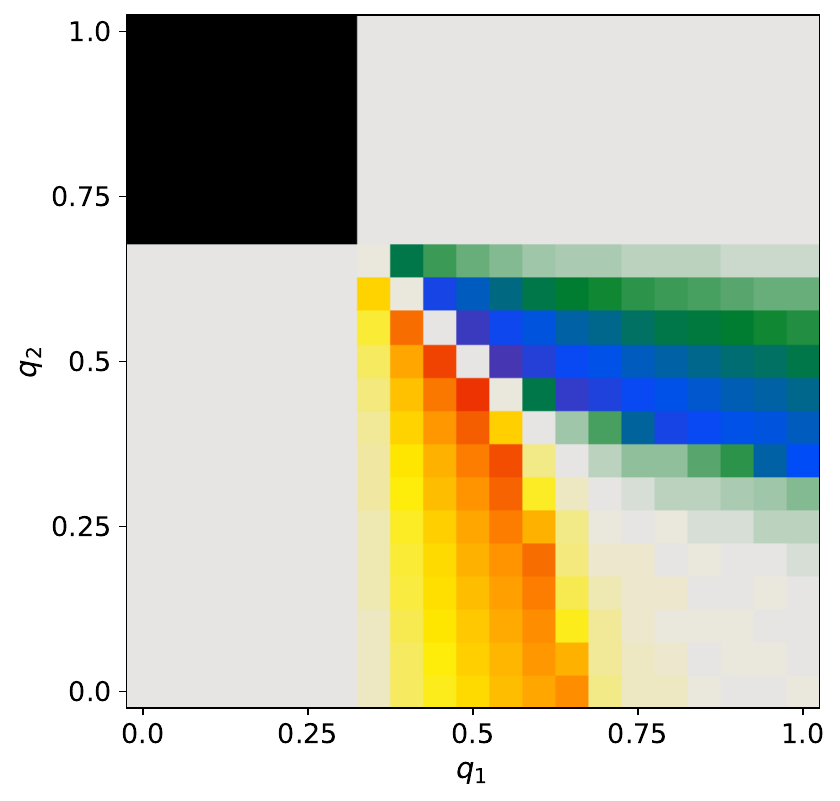}
\hskip 0.3 cm
\includegraphics[width=0.256\textwidth]{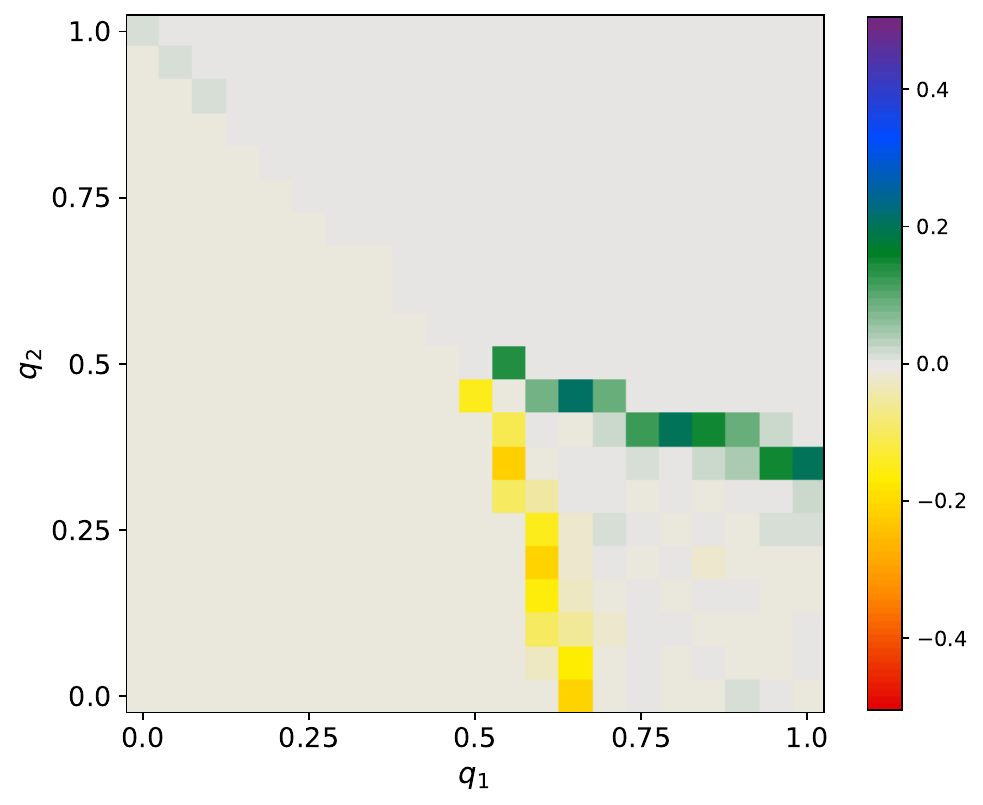}
    \caption{Difference between the stationary density measured with 
the Monte Carlo simulation and the Mean Field approximation on the 
left and 
    between the Monte Carlo simulation and $7$-block approximation 
on the right.
    As in Figure~\ref{fig:BBR_MC}
    the heatmap shows the measured quality for a given couple 
    of parameters $(q_1, q_2)$ on a grid of $21\times21$ points 
    in $[0,1]\times[0,1]$ with spacing $0.05$.
    In the black region on the left, the Mean Field solution is not acceptable. 
}
    \label{fig:BBR_confronto}
\end{figure}

In Figures~\ref{fig:BBR_MC_line_1} and \ref{fig:BBR_MC_line_2},
we have evaluated the effectiveness of the block approximation in describing the transitions observed via Monte Carlo simulations. 
Our findings confirm that the phase diagram of the model is 
highly non-trivial, exhibiting both continuous and 
discontinuous transition lines that converge at the point 
$(1/2,1/2)$.
As a final remark, we demonstrate that the block approximation 
can predict the stationary value of the density with acceptable 
precision across the entire parameter space 
$[0,1]\times[0,1]$. 
Additionally, we emphasize that it improves upon Mean Field results 
even when not excessively large maximal blocks are used.
Indeed, in Figure~\ref{fig:BBR_confronto}
we plot the difference between the stationary density 
measured with the Monte Carlo simulation and the Mean Field 
approximation on the left, and between the Monte Carlo simulation 
and the 7-block approximation on the right. 
As usual, the iterative equations are solved with a precision of $10^{-7}$.
The figure clearly shows that the block approximation is 
much more reliable than the Mean Field computation. Moreover, 
errors (which can be reduced by increasing the size of the maximal block) 
in estimating the stationary density are detectable only very 
close to the continuous transition lines.

\section{Conclusions}
\label{s:con} 
\par\noindent
This work aimed to compare the effectiveness of a crude implementation 
of block 
approximation with respect to other approaches, such 
as Monte Carlo and Mean Field, in describing the phase 
diagram of a \gls{pca}.
We studied several cases with increasingly complex dynamics, 
including the \gls{ndeca}, the diploid with local rules 200-182, 
and an example of totalistic probabilistic \gls{pca}.

Overall, the Mean Field approach, while computationally 
efficient and straightforward, often fails to capture the dynamical features and the transitions between different 
stationary states. 
In contrast, the block approximation, 
although more computationally intensive, always provides a 
more accurate and detailed representation for all \glspl{pca} considered. 
In some cases, successfully identifying transitions was possible even 
with a low-order block approximation.

Regarding \glspl{ndeca}, all results supported the ability of 
block approximation to accurately reproduce their dynamics compared 
to the Mean Field approach. For instance, in the case 
of \gls{ndeca} 202, the block approximation successfully reproduced 
the transition, a feature not captured by the Mean Field approach. 
Similarly, for \gls{ndeca} 234, the Mean Field approach predicted 
a phase transition not observed in Monte Carlo simulations. 
The block approximation, however, showed no phase transition and 
aligned well with Monte Carlo behaviour as larger blocks were 
included in the analysis.
The case of \gls{ndeca} 120 exemplified a scenario where the Mean Field 
predicted a phase transition that did not actually occur. Here, the 
block approximation revealed an apparent phase transition. However, 
upon inspecting the block probabilities, it became evident that this 
apparent transition was likely the result of artefacts induced by 
the approximation method. In the case of \gls{ndeca} 222, 
the block approximation closely approximated the transition 
point, matching the Monte Carlo results where the Mean Field approach 
diverged significantly.

Exploring the dynamics of the diploid 200-182 enabled us to examine 
the performance of the block approximation method, still in our 
raw version, in the context of a more intricate diploid, which does not 
include the null rule and in the presence of two absorbing states. 
Initially, we noted that the block approximation 
accurately replicated the behaviour of the active region near $\lambda=0$, 
a result unexpected by the Mean Field approach. Additionally, the block 
approximation provided interesting results in the region above 
the transition point, when the predictable structures of the local 182 
become slightly more relevant than those of the local rule 200. 
We also observed that, increasing the order of the approximation, 
a better estimate of the transition point and a better match 
with the Monte Carlo density profile was found. 
Applying the block approximation to the totalistic probabilistic \gls{pca} 
revealed that it can lead to very accurate descriptions. Unlike 
the Mean Field approach, which did not identify any transition along 
the line $q_2-q_1=-1/2$, the block approximation revealed the presence 
of a discontinuous transition. 
Specifically, the block approximation revealed to be able 
to capture the complicated structure of the phase diagram in 
which two continuous transitions are present. 
Overall, the analysis of the Mean Field and block approximation 
methods highlight the trade-off between computational feasibility 
and accuracy. The Mean Field approach, intended as a useful initial 
analysis tool can quickly provide insights into the general 
behavior of \glspl{pca} within a theoretical framework. However, 
our analysis emphasizes that this approach might lead to erroneous 
predictions, either missing transitions where they exist or 
predicting them where they do not.
A simple crude implementation of the 
3-block approximation offers a practical solution to 
outline the system phase diagram without involving complex 
and time-consuming computations. For a more precise and 
comprehensive understanding, particularly in systems exhibiting 
complex phase transitions, the block approximation proves 
to be a valuable tool. Combining both approaches allows for 
a balance of efficiency and accuracy, employing the Mean Field 
approach for broad theoretical overviews and the block approximation 
for detailed investigations.

\begin{acknowledgments}
ENMC and GL 
thank the PRIN 2022 project
``Mathematical Modelling of Heterogeneous Systems (MMHS)",
financed by the European Union - Next Generation EU,
CUP B53D23009360006, Project Code 2022MKB7MM, PNRR M4.C2.1.1.
\end{acknowledgments}

\begin{DeclarationInterest}
The authors have nothing to declare 
\end{DeclarationInterest}

\begin{DeclarationGenAI}
The authors declare that no generative AI and AI-assisted technologies have been utilized during the writing process of this manuscript.
\end{DeclarationGenAI}


\end{multicols}

\bibliographystyle{unsrt}

\bibliography{cls-pca_ba}

\begin{thebibliography}{10}

\bibitem{LN2016}
R.~Fernandez, P.-Y. Louis, and F.R. Nardi.
\newblock Overview: Pca models and issues.
\newblock In P.-Y Louis and F.~R. Nardi, editors, {\em Probabilistic Cellular Automata: theory, applications and future perspectives}, pages 1--30. Springer International Publishing, 2018.

\bibitem{VRD2023}
C.A. Valentim, J.A. Rabi, and S.A. David.
\newblock Cellular-automato model for tumor growth dynamics: virtualization of different scenarios.
\newblock {\em Computers in Biology and Medicine}, 153:106481, 2023.

\bibitem{D2020}
P.~Davis.
\newblock Does new physics lurk inside living matter?
\newblock {\em Physics Today}, 73(8):34, 2020.

\bibitem{W83}
S.~Wolfram.
\newblock Statistical mechanics of cellular automata.
\newblock {\em Rev. Mod. Phys.}, 35:601--644, 1983.

\bibitem{W84}
S.~Wolfram.
\newblock Computation theory of cellular automata.
\newblock {\em Comm. Math. Phys.}, 96:15--57, 1984.

\bibitem{Fautomata2017}
N.~Fat\'es.
\newblock Diploid celluar automata: First experiments on the random mixtures of two elementary rules.
\newblock {\em Lectures Notes in Computer Science}, 10248:97--108, 2017.

\bibitem{BMM2013}
A.~Busic, J.~Mairesse, and I.~Marcovici.
\newblock Probabilistic cellular automata, invariant measures, and perfect sampling.
\newblock {\em Adv. in Appl. Probab.}, 45:960--1980, 2013.

\bibitem{T2004}
A.~Toom.
\newblock {\em Contours, convex sets and cellular automata}.
\newblock IMPA mathematical publications, 2004.

\bibitem{MM2014}
J.~Mairesse and I.~Marcovici.
\newblock Around probabilistic cellular automata.
\newblock {\em Theoretical Computer Science}, 559:42--72, 2014.

\bibitem{Me2011}
J.R.G. Mendon\c{c}a.
\newblock Monte carlo investigation of the critical behavior of stavskaya's probabilistic cellular automaton.
\newblock {\em Phys. Rev. E}, 83:42--72, 2011.

\bibitem{Dh83}
D.~Dhar.
\newblock Exact solution of a directed-site animals-enumeration problem in three dimensions.
\newblock {\em Phys. Rev. Lett.}, 51:853--856, 1983.

\bibitem{GVK87}
H.A. Gutowitz, J.D. Victor, and B.W. Knight.
\newblock Local structure theory for cellular automata.
\newblock {\em Physica D}, 28:18--48, 1987.

\bibitem{GV87}
H.A. Gutowitz and J.D. Victor.
\newblock Local structure theory in more than one dimension.
\newblock {\em Complex Systems}, 1:57--68, 1987.

\bibitem{FF15}
H.~Fuk\'s and N.~Fat\`es.
\newblock Local structure approximation as a predictor of seconnd order phase transitions in asynchroous cellular automata.
\newblock {\em Natural computing}, 14:507--522, 2015.

\bibitem{F12}
H.~Fuk\'s.
\newblock Construction of local structure maps for cellular automata.
\newblock {\em J. Cell. Automata}, 7:455--488, 2012.

\bibitem{Me2011a}
J.R.G. Mendon\c{c}a and M.J. de~Oliveira.
\newblock An extinction-survival-type phase transition in the probabilistic cellular automaton p182--q200.
\newblock {\em Journal of Physics A: Mathematical and Theoretical}, 44:155001, 2011.

\bibitem{P05}
A.~Pelizzola.
\newblock Cluster variation method in statistcal physics and probabilstic graphical models.
\newblock {\em J. Phys. A: Math. Gen.}, 38:R309, 2005.

\bibitem{CGP1996}
E.N.M. Cirillo, G.~Gonnella, and A.~Pelizzola.
\newblock Folding transition of the triangular lattice in a discrete three--dimensional space.
\newblock {\em Physical Review E}, 53:3253, 1996.

\bibitem{CNS2021}
E.N.M. Cirillo, F.R. Nardi, and C.~Spitoni.
\newblock Phase transitions in random mixtures of elementary cellular automata.
\newblock {\em Physica A}, 573:125942, 2021.

\bibitem{MScmp1991}
C.~Maes and S.~Shlosman.
\newblock Ergodicity of probabilistic cellular automata: A constructive criterion.
\newblock {\em Communication in Mathematical Physics}, 135:233--251, 1991.

\bibitem{MScmp1993}
C.~Maes and S.~Shlosman.
\newblock When is an interacting particle system ergodic?
\newblock {\em Communication in Mathematical Physics}, 151:447--466, 1993.

\bibitem{Ba2000}
F.~Bagnoli, N.~Boccara, and R.~Rechtman.
\newblock Nature of phase transitions in a probabilistic cellular automaton with two absorbing states.
\newblock {\em Physical Review E}, 63:046116, 2001.

\bibitem{K74}
R.~Kikuchi.
\newblock Superposition approximation and natural iteration calculation in cluster--variation method.
\newblock {\em J. Chem. Phys.}, 60:1071--1080, 1974.

\bibitem{taggi}
L.~Taggi.
\newblock Critical probabilities and convergence time of percolation probbilistic cellular automata.
\newblock {\em Journal of Statistical Physics}, 159:853--892, 2015.

\bibitem{W94}
S.~Wolfram.
\newblock {\em Cellular automata and complexity: collected papers}.
\newblock crc Press, 2018.

\bibitem{Grassberger:1982}
P.~Grassberger.
\newblock New mechanism for deterministic diffusion.
\newblock {\em Physical Review A}, 28:3666, 1983.

\end{thebibliography}

\end{document}